\documentclass[12pt,a4paper]{iopart}
\pdfoutput=1

\usepackage{cite}
\usepackage{tabularx}

\usepackage{graphicx}
\begin{document}

\title{Overlap renormalization group transformations for disordered systems}

\author{Dimitrios Bachtis}

\address{Laboratoire de Physique de l'Ecole Normale Sup\'erieure, ENS, Universit\'e PSL,
CNRS, Sorbonne Universit\'e, Universit\'e de Paris, F-75005 Paris, France}
\ead{dimitrios.bachtis@phys.ens.fr}
\vspace{10pt}
\begin{indented}
\item[]August 2023
\end{indented}

\begin{abstract}
We establish a renormalization group approach which is implemented on the degrees of freedom defined by the overlap of two replicas to determine the critical fixed point and to extract four critical exponents for the phase transition of the three-dimensional Edwards-Anderson model. In addition, we couple the overlap order parameter to a fictitious field and introduce it within the two-replica Hamiltonian of the system to study its explicit symmetry-breaking with the renormalization group.   Overlap transformations do not require a renormalization of the random couplings of a system to extract the critical exponents associated with the relevant variables of the renormalization group. We conclude by discussing the applicability of such transformations in the study of any phase transition which is fully characterized by an overlap order parameter.
\end{abstract}

\vspace{2pc}

\section{\label{sec:level1}Introduction}

The renormalization group~\cite{PhysicsPhysiqueFizika.2.263,PhysRevB.4.3174,PhysRevLett.28.240,WILSON197475,RevModPhys.47.773} is a mathematical framework which has been traditionally utilized to advance our understanding of phase transitions within  statistical mechanics or quantum field theory. Nevertheless, traditional approaches of the renormalization group, which have been so widely successful in different subfields of physics, do not straightforwardly extend to all cases. An example is the theory of disordered systems. Within disordered systems and, specifically, spin glasses~\cite{Mezard1987} a novel set of problems emerge. 

The development of the renormalization group in disordered systems~\cite{Young_1976a,PhysRevLett.36.415,Southern_1977,Young1977b,PhysRevB.37.7745,PhysRevB.87.134201,PhysRevLett.114.095701,Angelini2017} is expected to necessitate the conception of a transformation which does not only transform the degrees of freedom themselves, but is also capable of handling the inherent stochasticity within the system. This inherent stochasticity, which must be considered as a renormalizable quantity, emerges from the random couplings: these are drawn from a predefined probability distribution and give rise to inhomogeneous interactions over the degrees of freedom. An additional renormalization over the probability distribution of the random couplings appears to become a necessary step and it emerges as a nontrivial problem.

From a practical perspective, the success of a renormalization group transformation is contingent on the flows that it induces within the system's parameter space: these can be observed in relation to the order parameter which fully characterizes the phase transition of the system. In the case of glassy systems, the correct order parameter is often given by the overlap~\cite{PhysRevLett.43.1754,PhysRevLett.50.1946,PhysRevLett.52.1156,Edwards_1975}. One must therefore confront the problem of devising a renormalization group transformation which preserves the probabilistic interpretation of the system and is also capable of properly transforming the overlap order parameter, a quantity defined over multiple replicas.

In this manuscript,  we implement renormalization group transformations on the degrees of freedom defined by the overlap of two replicas to determine the critical fixed point and to extract multiple critical exponents for the phase transition of the three-dimensional Edwards-Anderson model. The method is related to techniques which obtain insights from overlap configurations~\cite{PhysRevLett.55.2606,Lewenstein_2022,PhysRevB.98.174205,PhysRevB.98.174206}. Specifically, it is conceptually related to  the Haake-Lewenstein-Wilkens approach~\cite{PhysRevLett.55.2606,Lewenstein_2022} and provides a natural extension of the renormalization group in relation to the Parisi interpretation of the overlap order parameter and, explicitly, the overlap variables. 

To our knowledge, renormalization group transformations for spin glasses were first discussed in a computational setting by Young and Stinchcombe~\cite{Young_1976a}. The presence of a phase transition in the three-dimensional Edwards-Anderson model was then investigated with the renormalization group by Southern and Young~\cite{Southern_1977}.  Monte Carlo renormalization group methods were implemented by Wang and Swendsen~\cite{PhysRevB.37.7745}, with the use of the replica Monte Carlo method~\cite{PhysRevLett.57.2607}, and via calculations of the linearized renormalization group transformation matrix~\cite{PhysRevLett.42.859} to calculate two exponents. 

In the current manuscript, we do not utilize the linearized renormalization group transformation matrix and instead explore if we can conduct direct calculations of four exponents using the Wilson two-lattice matching Monte Carlo renormalization group method~\cite{Wilsonbook,PhysRevB.29.4030} and  a computational implementation of the Kadanoff scaling picture~\cite{PhysicsPhysiqueFizika.2.263}.  It is therefore possible to comment,  based on the obtained results, on the consistency of distinct renormalization group methods.  Furthermore, we utilize the renormalization group to study the explicit symmetry breaking within the two-replica Hamiltonian of the three-dimensional Edwards-Anderson model. Specifically, we couple the overlap order parameter to a fictitious field and introduce it within the two-replica Hamiltonian to break explicitly its symmetry. We then construct mappings which relate the original and the renormalized fictitious fields, and extract the critical exponent associated with the divergence of the correlation length in relation to the field of the overlap order parameter. The computational implementation is based on the parallel tempering method~\cite{doi:10.1143/JPSJ.65.1604,Marinari_1992} and the introduction of reweighting techniques for renormalized observables~\cite{bachtis2020adding,PhysRevLett.128.081603,arxiv.2205.08156} to spin glasses, via the multiple histogram method~\cite{PhysRevLett.63.1195}.

 We remark that the current study is a proof-of-principle demonstration that is orders of magnitude smaller in computational effort than prior large-scale simulations or studies conducted with the use of supercomputing facilities, and it is not meant to compete in numerical precision with those. However, we discuss potential benefits of Monte Carlo renormalization group methods in the study of spin glasses. These include the partial elimination of finite-size effects, the linearization of the transformation which enables accurate calculations of critical exponents anywhere in the linear region of the fixed point, and the fact that configurations of smaller lattice sizes can be constructed by the application of the renormalization group transformation: all computational resources could then be employed to sample the largest lattice size possible within the context of a Monte Carlo renormalization group study and the smaller lattice sizes could be obtained via the application of the transformation, and without requiring additional Monte Carlo simulations. Finally, we conclude by discussing how overlap renormalization group transformations, which do not require a renormalization of the random couplings of a system, enable the extraction of multiple critical exponents in any phase transition which is fully characterized by an overlap order parameter.

\begin{figure*}[t]
\begin{indented}
\item[]\includegraphics[width=12.2cm]{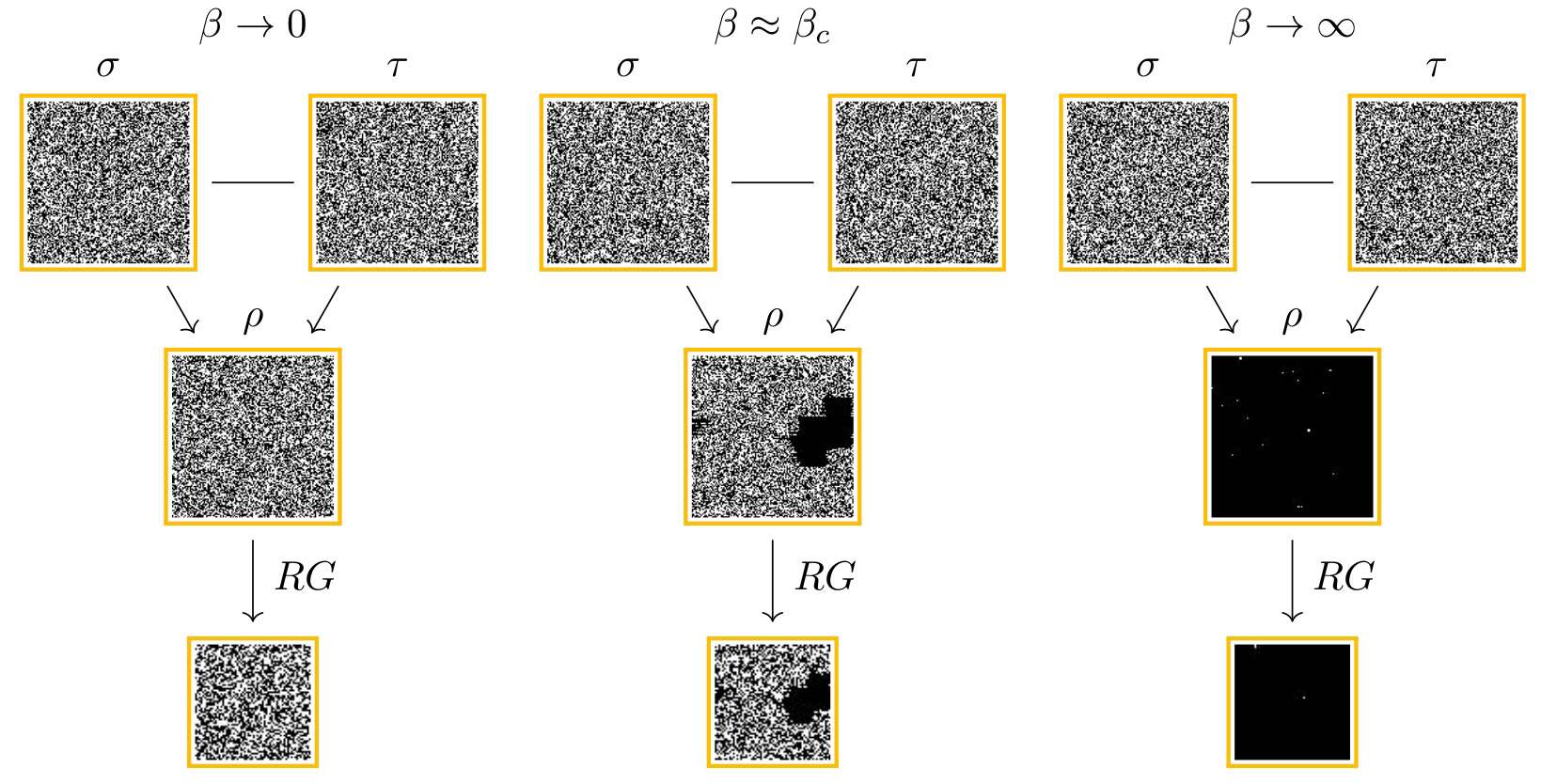}
 \caption{\label{fig:rg} Illustrative examples of the renormalization group implementation based on the mapping of two replicas $\sigma$, $\tau$ to an overlap configuration $\rho$. The spin glass phase transition of the three-dimensional Edwards-Anderson model is mapped to a system with overlap degrees of freedom which manifests critical behavior that resembles ferromagnetism. The emergence of the spin glass correlation length can then be explicitly observed in the overlap configuration based on the creation of clusters. Renormalization group transformations, using the majority rule, are directly implemented on the overlap degrees of freedom. The $\beta \rightarrow \infty$ illustration depicts the case where the overlap order parameter $q\approx-1$. }
\end{indented}
\end{figure*}

\section{Spin glasses and the overlap}

We consider the case of the three-dimensional Edwards-Anderson model with three replicas $\sigma$, $\tau$, $\upsilon$ which comprise spins $s$, $t$, $u$, respectively. To simplify notation we will derive equations based on two replicas $\sigma$, $\tau$. For a specific inverse temperature $\beta$ the Hamiltonian of the system is defined as:
\begin{equation}\label{eq:origham}
E_{\sigma,\tau} = E_{\sigma}+E_{\tau}=-\sum_{\langle ij \rangle} J_{ij} (s_{i}s_{j}+t_{i}t_{j}),
\end{equation}
where $\langle ij \rangle$ denotes a nearest-neighbor interaction for spins $s,t=\pm1$, and $J_{ij}$ are random couplings sampled as $J_{ij}=\pm1$ with equal probability. The replicas experience the same realization of disorder $\lbrace J_{ij} \rbrace$. 

The system defines a joint Boltzmann probability distribution $p_{\sigma_{i},\tau_{j}}$ for two configurations $\sigma_{i}$, $\tau_{j}$ at inverse temperature $\beta$:
\begin{equation}\label{eq:origprob}
p_{\sigma_{i},\tau_{j}}= \frac{\exp[-\beta(E_{\sigma_{i}}+ E_{\tau_{j}})]}{\sum_{\sigma}\sum_{\tau} \exp[-\beta(E_{\sigma}+ E_{\tau})]},
\end{equation}
where $Z_{\sigma,\tau}=\sum_{\sigma}\sum_{\tau} \exp[-\beta(E_{\sigma}+ E_{\tau})]$ is the partition function and the sums are over all possible configurations $\sigma$, $\tau$. We remark that $Z_{\sigma,\tau}=Z_{\sigma}Z_{\tau}=Z_{\sigma}^{2}=Z_{\tau}^{2} \equiv Z^{2}$ in terms of the equilibrium occupation probabilities. However we keep the partition functions separated as we aim to estimate them independently based on a finite size of samples which is obtained from Markov chain Monte Carlo simulations.

The phase transition of the three-dimensional Edwards-Anderson model is characterized by the overlap order parameter $q$, defined in terms of two replicas  $\sigma$ and $\tau$:
\begin{equation}
q_{\sigma\tau}= \frac{1}{V} \sum_{i} s_{i} t_{i},
\end{equation} 
where $V=L^{3}$ is the volume of the system.

 The overlap order parameter is bound between $[-1,1]$. When in the spin glass phase, the value $q=1$ corresponds to the case where the spins in each set $\lbrace s_{i},t_{i} \rbrace$ of the two replicas $\sigma$, $\tau$ have identical values. In constrast, the value $q=-1$ corresponds to the case where all the spins $\lbrace s_{i},t_{i} \rbrace$  in the two replicas have exactly different values. In addition one can define the overlap susceptibility $\chi_{q}$ as:
\begin{equation}
\chi_{q}=\beta V(\langle q^{2} \rangle-\langle q \rangle^{2})= \beta V \langle q^{2}  \rangle,
\end{equation} 
where $\langle q \rangle=0$.

It is now possible to map the two-replica system, based on the Haake-Lewenstein-Wilkens perspective, into a single system which comprises overlap degrees of freedom. Specifically, given two replicas $\sigma$, $\tau$ with spins $s$, $t$ and lattice size $L$ in each dimension  we define a configuration $\rho$ with degrees of freedom $\varrho_{i}=s_{i} t_{i}$ and  lattice size $\mathcal{L}=L$. This mapping defines an effective Hamiltonian $E^{\textrm{eff}}$, partition function $Z^{\textrm{eff}}$, and Boltzmann probability distribution $p^{\textrm{eff}}_{\rho_{i}}$ for the three-dimensional Edwards-Anderson model. The effective probability distribution  $p^{\textrm{eff}}_{\rho_{i}}$, averaged over disorder,  has the form~\cite{PhysRevLett.55.2606}:

\begin{eqnarray*}
p^{\textrm{eff}}_{\rho_{i}}   & = & \frac{\exp\big[-\beta E^{\textrm{eff}}_{\rho_{i}} \big]}{Z^{\textrm{eff}}_{\rho}} \\ & = & \Bigg[ \sum_{\lbrace s \rbrace} \frac{\exp\big[\beta \sum_{\langle ij \rangle} J_{ij}(1+\varrho_{i}\varrho_{j})s_{i} s_{j}\big]}{Z^{2}[\lbrace J_{ij} \rbrace]}\Bigg]_{J_{ij}} \\ & = & 2^{V} \Bigg[   \frac{\exp\big[\beta \sum_{\langle ij \rangle} J_{ij}(1+\varrho_{i}\varrho_{j})\big]}{Z^{2}[\lbrace J_{ij} \rbrace]} \Bigg]_{J_{ij}},
\end{eqnarray*}

where the gauge transformation $J_{ij} \rightarrow J_{ij}s_{i}s_{j}$ was used to obtain the last equation.

The first implication, which emerges from the aforementioned mapping, is that one can define the equivalent of a magnetization $m$ in the system with overlap degrees of freedom $\varrho$:
\begin{equation}
m=\frac{1}{V} \sum_{i} \varrho_{i}.
\end{equation}

By recalling that $\varrho_{i}=s_{i}t_{i}$ we establish that the spin glass phase transition of the two-replica Hamiltonian, which is characterized by the overlap order parameter $q$, is now mapped to the phase transition of an effective system, characterized by the magnetization $m$, and with critical behavior which resembles ferromagnetism.  Formally:
\begin{equation}
[\langle q \rangle]=\Big[  \sum_{\sigma} \sum_{\tau} q_{\sigma \tau} p_{\sigma,\tau} \Big]=  \sum_{\rho} m_{\rho} p_{\rho}^{\textrm{eff}} = \langle m \rangle_{\textrm{eff}},
\end{equation}
where $\langle \rangle$ is the thermal average and $[ ]$ the average over realizations of disorder.

The mapping from the two-replica system to the effective system has certain implications in relation to the phase transition. Specifically, if we denote as $\xi(\sigma,\tau,\beta)$ and $\xi(\rho,\beta)$ the correlation lengths of the two-replica and the effective system respectively, then  $\xi(\sigma,\tau,\beta)=\xi(\rho,\beta) \ \forall  \ \beta \in (0,\infty)$. The two systems therefore have an identical correlation length for all inverse temperatures $\beta$. Since the correlation length diverges at the critical fixed point $\xi(\sigma,\tau,\beta_{c},L=\infty)=\xi(\rho,\beta_{c},\mathcal{L}=\infty)=\infty$ the aforementioned observation implies that the two systems have an identical fixed point $\beta_{c}$. In addition since observables of the two systems diverge according to the same correlation length $\xi$ this implies that the critical exponents of quantities associated with the overlap order parameter $q$ or, equivalently, with the magnetization $m$ are also identical for the two systems.

In this manuscript we are interested in implementing renormalization group transformations on the overlap degrees of freedom $\varrho$. To achieve this, one must formally apply the renormalization group on a probability distribution which is a function of exclusively the overlap variables $\varrho$. To rephrase, the implementation of a renormalization group transformation on the overlap variables $\varrho$ necessitates the development of an effective theory for the overlap system. The Haake-Lewenstein-Wilkens approach provides such an effective theory by introducing a system which is described by an effective Hamiltonian $E^{\textrm{eff}}(\rho)$, partition function $Z^{\textrm{eff}}(\rho)$, and Boltzmann probability distribution $p^{\textrm{eff}}_{\rho_{i}}$, all of which are functions of exclusively the overlap variables $\varrho$. Equivalently, the Haake-Lewenstein-Wilkens approach enables a mathematically formal implementation of the renormalization group on the effective overlap system. We remark that it remains possible that other theories can emerge to describe the overlap system. 

Monte Carlo renormalization group methods were first developed for the effective system by Wang and Swendsen~\cite{PhysRevB.37.7745}.  One considers that the application of a renormalization group transformation is established on the effective probability distribution $p^{\textrm{eff}}$ to obtain a renormalized probability distribution $p^{'\textrm{eff}}$ as:
\begin{equation}
p^{'\textrm{eff}}_{\rho'}= \sum_{\rho} T(\rho',\rho) p^{\textrm{eff}}_{\rho}, 
\end{equation}
where $T(\rho',\rho)$ corresponds to the majority rule. Since $T(\rho',\rho)$ has no explicit dependence on the random couplings the renormalization group transformation can be interchanged with the averaging over the realizations of disorder. The approach, including the application of renormalization group transformations on the system with overlap degrees of freedom is illustrated in Fig.~\ref{fig:rg}. 

We remark that, while it is of interest to introduce, within a renormalization group setting, correlation functions which have, for instance, the form of neighbor interactions $\varrho_{i} \varrho_{j}$ between the overlap degrees of freedom $\varrho$, we will not pursue this direction here. Instead, we will focus solely on quantities derived from the overlap order parameter $q$. These remain, probabilistically, observables of the original system and are therefore Boltzmann-distributed based on the original probability distribution of~(\ref{eq:origprob}). We will then incorporate reweighting techniques within the computational renormalization group method to determine the critical fixed point and to extract four critical exponents. 

\section{\label{sec:level2}Parallel Tempering and Multi-histogram Reweighting}

 While one can, in principle, simulate directly the three-dimensional Edwards-Anderson model using a conventional Markov chain Monte Carlo simulation with the Metropolis algorithm, such computations can be prohibitively long due to issues pertinent to the thermalization of the Markov chain and the critical slowing down effect. To alleviate simulational problems, we utilize the parallel tempering technique. 
 
 We consider a set of inverse temperatures  $\lbrace \beta_{0}, \beta_{1},\ldots, \beta_{m} \rbrace$, with $\beta_{0}<\beta_{1}<\beta_{m}$. Specifically, we consider $m=22$ and $\beta_{0}=0.55$, $\beta_{m}=1.15$ where the set of inverse temperatures extends to the spin glass phase. In addition we impose the constraint that two adjacent inverse temperatures are sufficiently close in parameter space: this implies that there exists an overlap between the energy histograms for systems simulated at the two inverse temperatures. 

We implement the parallel tempering technique by attempting an exchange of configurations for two simulated replicas at adjacent inverse temperatures. Specifically, if we consider two replicas at inverse temperatures $\beta_{0}$ and $\beta_{1}$ which have an energy difference of $\Delta E= E_{1}-E_{0}$, then an exchange of configurations is accepted based on the acceptance probability $\mathcal{A}$:
\begin{equation}
  \mathcal{A} =
    \cases{
      \exp [ -(\beta_{0}-\beta_{1}) \Delta E ]  &  if \ $\Delta E > 0$ \cr
       1 &  otherwise. \cr
           }
\end{equation}

We are additionally interested in a complementary method, namely the multi-histogram reweighting technique. We aim to utilize the multiple histogram method to reweight observables of the renormalized systems under the probability distributions of the original systems. To the best of our knowledge, while reweighting has been utilized in disordered systems~\cite{Janke2003},  this manuscript documents the first use of the multiple histogram method within a Monte Carlo renormalization group calculation of a disordered system.  Detailed derivations for the multiple histogram method can be found, for instance, in~\cite{newmanb99,PhysRevE.102.053306}.

To implement the multi-histogram reweighting technique for the replica $\sigma$ we optimize the partition function $Z_{\sigma,l}$ for each inverse temperature which belongs in the set $\lbrace \beta_{0}, \beta_{1},\ldots, \beta_{m} \rbrace$ via:
\begin{equation}
Z_{\sigma,l}= \sum_{i,x} \frac{1}{\sum_{j}n_{j}Z_{\sigma,j}^{-1}\exp[(\beta_{l}-\beta_{j})E_{ix}]},
\end{equation}
where the sum is over configurations $x$ which are sampled at simulation $i$ and have energy $E_{ix}$,  $j$ is a sum over all simulations at inverse temperatures of the set $\lbrace \beta_{0}, \beta_{1},\ldots, \beta_{m} \rbrace$ and $n_{j}$ is the number of configurations sampled at simulation $j$. After convergence is achieved, we estimate the partition function $Z_{\sigma}(\beta)$ for every inverse temperature $\beta$ in the continuous range $\beta \in [\beta_{0},\beta_{m}]$ via:
\begin{equation}
Z_{\sigma}(\beta)= \sum_{i,x} \frac{1}{\sum_{j}n_{j}Z_{\sigma,j}^{-1}\exp[(\beta-\beta_{j})E_{ix}]}.
\end{equation}

We complete the above process for the partition functions of both replicas $Z_{\sigma}$ and $Z_{\tau}$. For a given realization of disorder $J_{ij}$ the expectation value of any arbitrary observable $O$ defined on the two-replica Hamiltonian, can then be estimated via:

\begin{equation}
\langle O \rangle_{\beta} = \frac{1}{Z_{\sigma}(\beta)Z_{\tau}(\beta)} \sum_{i,x,y} \frac{O_{ixy}}{\sum_{j} n_{j} Z_{\sigma,j}^{-1} Z_{\tau,j}^{-1} \exp[-(\beta-\beta_{j})(E_{ix}+E_{iy})]} ,
\end{equation} 

where the sum is over configurations $x$, $y$ of replicas $\sigma$, $\tau$. We emphasize that the above equation can be additionally utilized to reweight observables $O'$ of renormalized systems under the original probability distribution by substituting every occurence of $O$ with $O'$. We consider an averaging over $100$ and $500$ realizations of disorder for systems with lattice sizes $L=16$ and $L=8$ in each dimension, respectively. The number of Monte Carlo samples used is in the range $10^{4}$ to $10^{5}$. The simulation details are summarized in Table~\ref{tab:simtab}.

 \begin{table}[t]
\begin{indented}
\setlength{\tabcolsep}{2.5em} 
\item[] \begin{tabular}{cccc}
\br
$L$& $N_{J_{ij}}$ & $N_{\beta}$  & $N_{\textrm{MC}}$ \\
\hline
$8$& $500$ & $22$  & $10^{4}$-$10^{5}$  \\
$16$ & $100$ & $22$ & $10^{4}$-$10^{5}$ \\
\br

\end{tabular}
\end{indented}
\caption{\label{tab:simtab} The simulation details where $N_{J_{ij}}$ corresponds to the number of the realizations of disorder, $N_{\beta}$ is the number of simulated inverse temperatures in the parallel tempering implementation which reside in the range $\beta \in [0.55,1.15]$, and $N_{\textrm{MC}}$ is the number of Monte Carlo steps conducted after the thermalization of the system was achieved based on the criteria of Bhatt and Young, see text for more details.  }
\end{table}

\begin{figure}[t!]
\begin{indented}
\item[]\includegraphics[width=8.6cm]{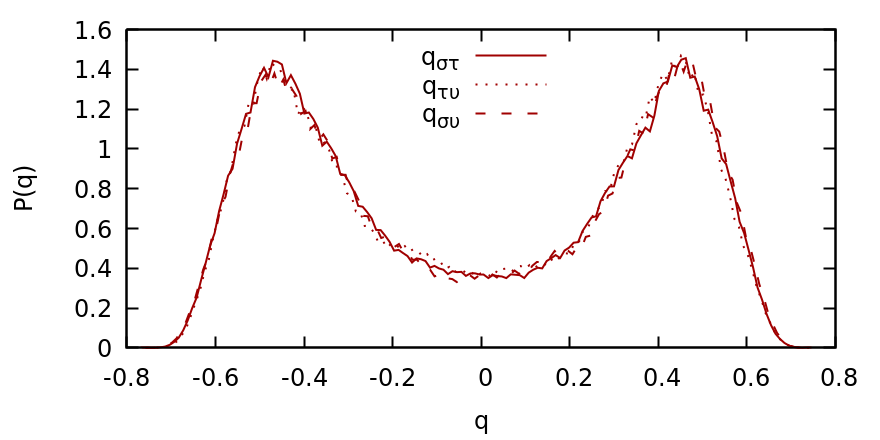}
\caption{\label{fig:fighist} The probability density function for the overlap order parameters $q_{\sigma\tau}$, $q_{\tau \upsilon}$, $q_{\sigma \upsilon}$ of three replicas $\sigma$, $\tau$, $\upsilon$ at inverse temperature $\beta=0.95$ and lattice size $L=16$ in each dimension. An initial number of configurations is discarded due to the thermalization of the Markov chain and the results are obtained based on subsequent measurements.}
\end{indented}
\end{figure}

In this manuscript we consider three replicas and thus we are able to calculate three values $q_{\sigma\tau}$, $q_{\tau \upsilon}$, $q_{\sigma \upsilon}$ of the overlap order parameter. In fact, we utilize these separate calculations, obtained via parallel tempering, to establish, based on the criteria proposed by Bhatt and Young~\cite{PhysRevLett.54.924}, that the ensemble has successfully thermalized. The probability density function for $q_{\sigma\tau}$, $q_{\tau \upsilon}$, $q_{\sigma \upsilon}$ of a system at inverse temperature $\beta=0.95$ with lattice size $L=16$ and one realization of disorder is depicted in Fig.~\ref{fig:fighist}. We observe that the distribution is symmetric and in approximate agreement for the three calculations, thus indicating that the system has successfully reached thermal equilibrium. 

We now utilize multi-histogram reweighting to interpolate the overlap order parameter in the range of $\beta \in [\beta_{0}, \beta_{m}]$. Results within the critical region are depicted in Fig.~\ref{fig:figorig}. We remark that multi-histogram reweighting is a technique which provides results in the entire critical region and simultaneously reduces the statistical errors. The method could then potentially prove useful in accurately locating the critical fixed point within the context of large-scale simulations for spin glasses.

\section{\label{sec:level3}Overlap renormalization group transformations}

\subsection{The renormalization group}

We are now able to implement a renormalization group transformation. which transforms the degrees of freedom defined by the overlap of the two replicas, for the three-dimensional Edwards-Anderson model. We recall that, given two replicas $\sigma$, $\tau$, we define a new lattice $\mathcal{L}$ where each site $i$ has the value $\varrho_{i}=s_{i} t_{i}$. Based on a rescaling factor of $b=2$ we then separate the lattice into blocks of size $b$ in each dimension and select the rescaled degree of freedom by applying the majority rule on the spins $\varrho$ within each block. When the number of positive and negative spins is equal we select randomly the rescaled degree of freedom.

\begin{figure}[t!]
\begin{indented}
\item[] \includegraphics[width=8.6cm]{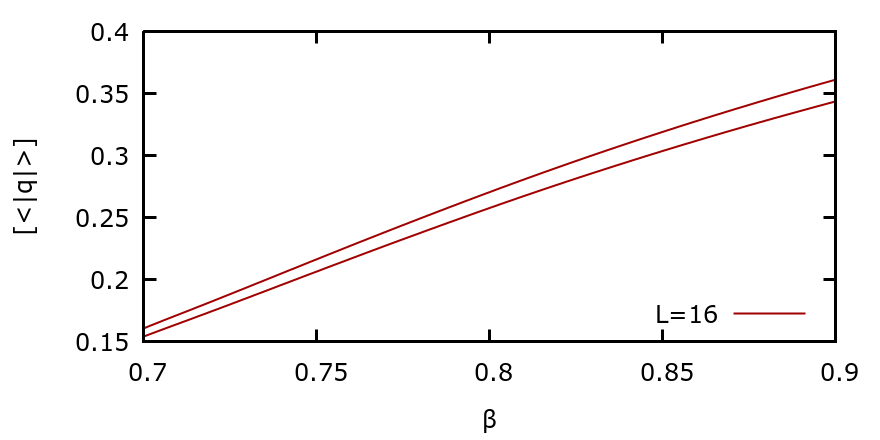}
\caption{\label{fig:figorig} Expectation of the absolute value of the overlap order parameter $q$ calculated under a thermal $\langle \rangle$ and disorder $[]$ average versus the inverse temperature $\beta$. The results are obtained using parallel tempering and multi-histogram reweighting on a two-replica Hamiltonian. The space bounded by the lines corresponds to the statistical uncertainty.}
\end{indented}
\end{figure}

Given an original system with overlap degrees of freedom and lattice size $\mathcal{L}$ or, equivalently, two replicas of lattice size $L$ in each dimension, we obtain via the renormalization group a rescaled system with lattice size $\mathcal{L}'$:
\begin{equation}
\mathcal{L}'=\frac{\mathcal{L}}{b}.
\end{equation}

The spin glass correlation length $\xi$ of the original system is therefore transformed,  in terms of lattice units, as:
\begin{equation}
\xi'=\frac{\xi}{b}.
\end{equation}

We remark that since the correlation length is a quantity which arises dynamically as we approach the critical fixed point, and the original and the renormalized systems are described by different correlation lengths $\xi$ and $\xi'$ then this implies that the two systems have a different distance from the critical fixed point. Consequently, the two systems are described by different inverse temperatures $\beta$ and $\beta'$ or, equivalently, by a different set of coupling constants. 

\subsection{The fixed point and the exponents $\gamma/\nu$, $\kappa_{q}/\nu$}

\begin{figure}[t]
\begin{indented}
\item[]\includegraphics[width=8.6cm]{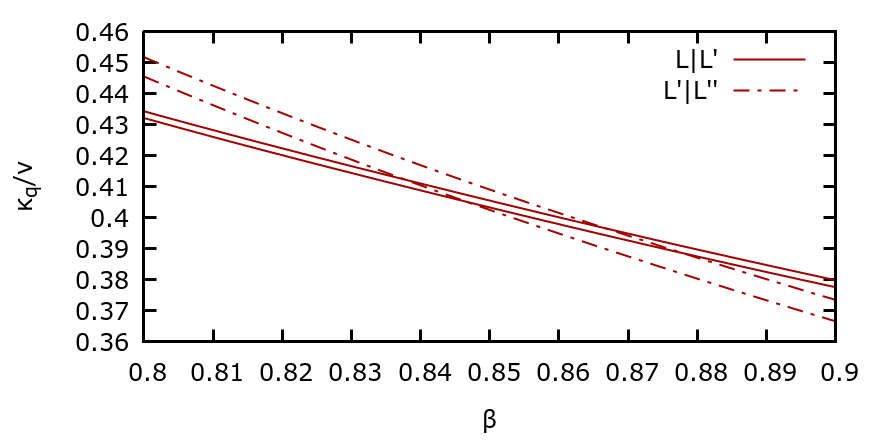}
\caption{\label{fig:figexp} The critical exponent $\kappa_{q} / \nu$ versus the inverse temperature $\beta$. The space bounded by the lines corresponds to the statistical uncertainty.}
\end{indented}
\end{figure}

\begin{figure}[t]
\begin{indented}
\item[] \includegraphics[width=8.6cm]{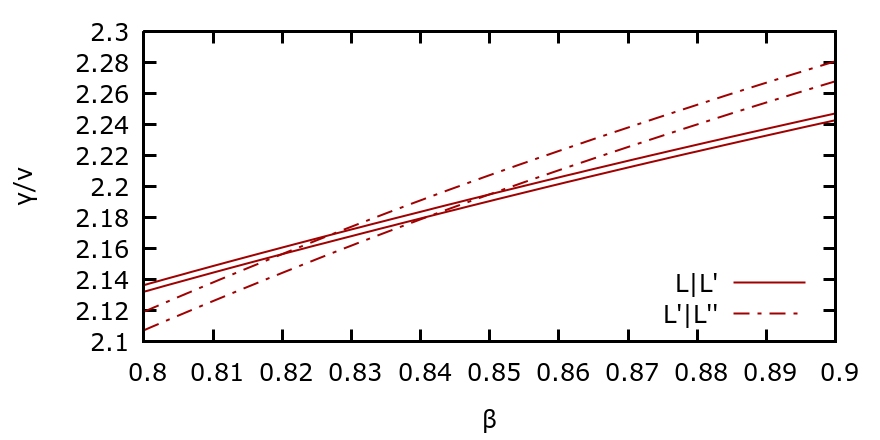}
\caption{\label{fig:figexp2} The critical exponent $\gamma / \nu$ versus the inverse temperature $\beta$. The space bounded by the lines corresponds to the statistical uncertainty.}
\end{indented}
\end{figure}

We are now able to conduct a renormalization group study of the three-dimensional Edwards-Anderson model based on transformations applied on the degrees of freedom $\varrho$ defined by the overlap of two replicas. This is achieved by utilizing exclusively quantities derived from the overlap order parameter $q$, thus avoiding the need to undergo a renormalization of the random couplings of the system. We will calculate the critical fixed point and two critical exponents via a direct computational implementation of the Kadanoff scaling picture.

As an initial step we are interested in locating the critical fixed point which describes the finite-temperature spin glass transition of the system at a critical $\beta_{c}$. We define the reduced coupling constants $t$ and $t'$ which measure the distance from the critical fixed point for an original and a renormalized system:
\begin{equation}
t=\frac{\beta_{c}-\beta}{\beta_{c}}, \ t'=\frac{\beta_{c}-\beta'}{\beta_{c}},
\end{equation}

We now define a critical exponent $\kappa_{q}$ which governs the divergence of the overlap order parameters $q$, $q'$ of the original and the renormalized systems:
\begin{equation}
[\langle |q| \rangle] \sim |t|^{\kappa_{q}}, \ [\langle |q'| \rangle] \sim |t'|^{\kappa_{q}}.
\end{equation}

We denote as $[\langle |q| \rangle]$ the absolute value of the overlap order parameter as calculated under a thermal $\langle \rangle$ and disorder $[]$ average.  We remark that while the exponent coupled to the order parameter is often denoted as $\beta$, we utilize $\kappa_{q}$ to avoid confusion with the inverse temperature $\beta$. The above equations can be equivalently expressed in terms of the spin glass correlation lengths $\xi$, $\xi'$ as:
\begin{equation}
[\langle |q| \rangle] \sim \xi^{-\kappa_{q}/\nu}, \ [\langle |q'| \rangle] \sim \xi'^{-\kappa_{q}/\nu},
\end{equation}
where $\nu$ is the correlation length exponent. By dividing the above equations and taking the natural logarithm we arrive at the expression:
\begin{equation}
\frac{\kappa_{q}}{\nu}= \frac{\log \frac{[\langle |q'| \rangle]}{[\langle |q| \rangle]}}{\log b}.
\end{equation}

Following an analogous process we are able to calculate the exponent $\gamma$ of the overlap susceptibility $\chi_{q}$ via:
\begin{equation}
\frac{\gamma}{\nu}= -\frac{\log \frac{[\langle \chi_{q}' \rangle]}{[\langle \chi_{q}\rangle]}}{\log b}.
\end{equation} 

We remark that since the correlation length diverges at the critical point $\xi(\beta_{c})=\xi'(\beta_{c})=\infty$, one anticipates that all intensive observables $O$, $O'$ of an original and a renormalized system intersect $O(\beta_{c})=O'(\beta_{c})$. Nevertheless, while this observation is mathematically valid, in practical implementations such intersections are subject to finite size-effects and are not always observed. However, there exists another approach to determine the fixed point even under the presence of finite-size effects: one can directly search for an intersection between the exponents themselves~\cite{PhysRevLett.43.177}. 

Based on the above observation, we utilize multi-histogram reweighting to obtain calculations of effective critical exponents $\kappa_{q}/\nu$, $\gamma/ \nu$ in the continuous range $\beta \in [\beta_{0},\beta_{m}]$. This simplifies the computational calculations since it is not necessary to conduct additional simulations. The results, obtained by two iterative renormalization group applications on a system with lattice size $\mathcal{L}=16$ to produce two renormalized systems of lattice sizes $\mathcal{L}_{16}'=8$ and $\mathcal{L}_{16}''=4$ in each dimension, are depicted in Figs.~\ref{fig:figexp} and~\ref{fig:figexp2}. We observe an intersection in parameter space for both calculations of the critical exponents. Consequently we obtain two estimates of the critical exponents $\gamma/\nu=2.15(2)$, $\kappa_{q}/\nu=0.40(1)$.

We remark that, in the previous renormalization group approach, calculations are conducted based on quantities derived from the overlap order parameter, which can be calculated on original and renormalized configurations without knowledge of the random couplings. In addition, renormalized observables $O'$ are reweighted under the original probability distribution and, therefore, under the original random couplings $\lbrace J_{ij} \rbrace $, which are known. In the next subsections, we will further improve on the previous calculations of the critical exponents by linearizing the renormalization group transformation.

\subsection{Inverse mappings and the exponent $\nu$}

We will now utilize a different approach, namely the Wilson two-lattice matching Monte Carlo renormalization group. The method concerns the comparison of an original and a renormalized system of identical lattice size $\mathcal{L}=\mathcal{L}'$. These two systems are described by different realizations of disorder $\lbrace J_{ij} \rbrace \neq \lbrace J_{ij}' \rbrace$. Consequently the renormalization group method to be discussed below is affected more strongly by systematic errors pertinent to the averaging over a small number of disorder realizations but is affected less by systematic errors pertinent to finite-size effects. 

To describe the method we consider again an original system of lattice size $\mathcal{L}=16$ to which we apply one renormalization group transformation to produce a renormalized system of lattice size $\mathcal{L}'_{16}=8$. We now compare observables of the renormalized system  $\mathcal{L}'_{16}=8$ against observables of an original system $\mathcal{L}$ with $\mathcal{L}=\mathcal{L}'_{16}=8$. By applying another renormalization group transformation on both systems we additionally conduct calculations for systems with $\mathcal{L}'_{8}=\mathcal{L}''_{16}=4$.  The benefit of the discussed renormalization group method is that by comparing two systems of identical lattice size one partially eliminates finite-size effects, and thus accurate calculations can be obtained on trivially small lattice sizes.

\begin{figure}[t]
\begin{indented}
\item[]\includegraphics[width=8.6cm]{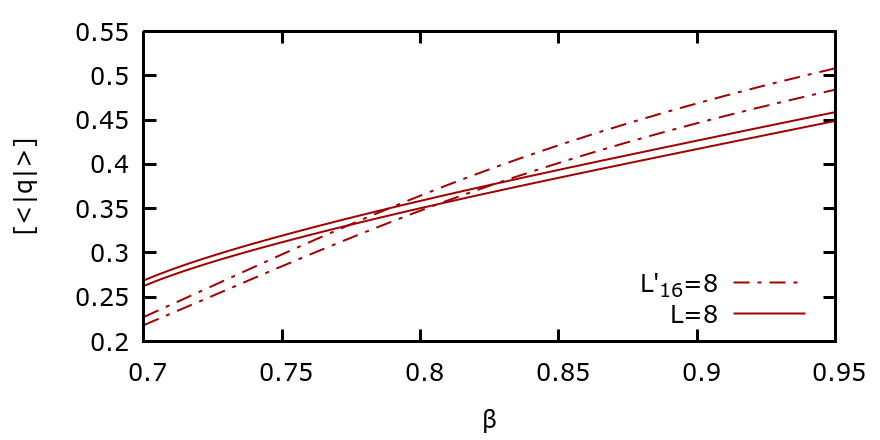}
\caption{\label{fig:figlm} Expectation of the absolute value of the overlap order parameter, calculated with multi-histogram reweighting under thermal and disorder averaging, for an original and a renormalized system of identical lattice size $L=L_{16}'=8$. The space bounded by the lines corresponds to the statistical uncertainty.}
\end{indented}
\end{figure}
\begin{figure}[t]
\begin{indented}
\item[]\includegraphics[width=8.6cm]{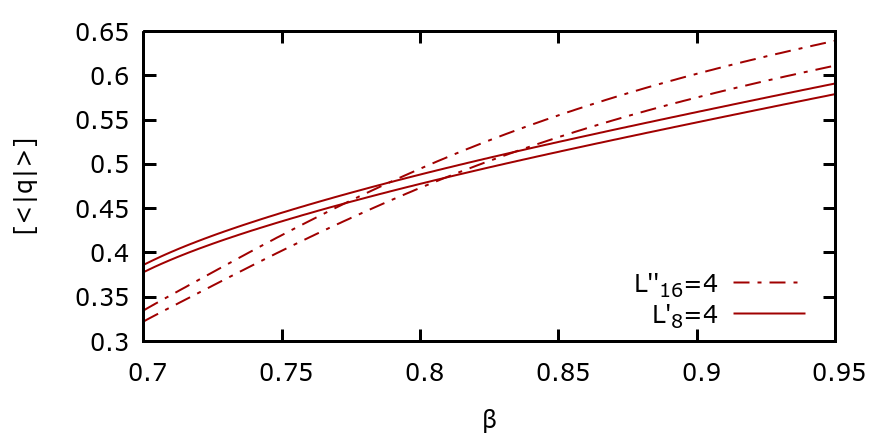}
\caption{\label{fig:figlm2} Expectation of the absolute value of the overlap order parameter, calculated with multi-histogram reweighting under thermal and disorder averaging, for two renormalized systems of identical lattice size $L_{8}'=L_{16}''=4$. The space bounded by the lines corresponds to the statistical uncertainty.}
\end{indented}
\end{figure}

We now utilize multi-histogram reweighting to obtain the absolute value of the overlap order parameter for systems with identical lattice size. The results are depicted in Figs.~\ref{fig:figlm} and \ref{fig:figlm2}. Under the reduction of finite-size effects we are now able to observe an intersection of observables at the critical fixed point. We additionally observe that the expected renormalization group flows have emerged. Specifically, for $\beta>\beta_{c}$ the overlap order parameter has values $[\langle |q'| \rangle] >[\langle |q| \rangle]$ since the system has been driven deeper to the spin glass phase due to the reduction of the correlation length. In an analogous manner, for $\beta<\beta_{c}$ the renormalized system has been driven towards the zero inverse temperature, and thus $[\langle |q'| \rangle] <[ \langle |q| \rangle]$.

Having established that the renormalization group method produces the anticipated behaviour, it is now possible to determine the renormalized coupling parameters. Specifically, we are interested in extracting a function which is able to relate, for instance, the original and the renormalized inverse temperatures $\beta$, $\beta'$. This can be achieved via the inverse mapping:
\begin{equation}
\beta'=O^{-1}(O'(\beta)).
\end{equation}   

\begin{figure}[t]
\begin{indented}
\item[]\includegraphics[width=8.6cm]{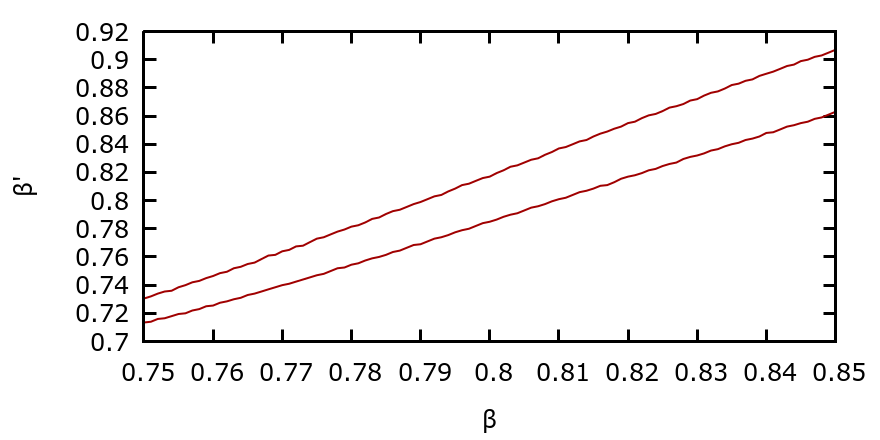}
\caption{\label{fig:figmapq} Renormalized inverse temperature $\beta'$ versus original inverse temperature $\beta$. The space bounded by the lines corresponds to the statistical uncertainty. }
\end{indented}
\end{figure}

 We can now utilize these mappings to extract the correlation length exponent $\nu$. Specifically, the divergence of the correlation lengths $\xi$, $\xi'$ in terms of the reduced inverse temperatures $t$, $t'$ is given by:
\begin{equation}
\xi \sim |t|^{-\nu} , \ \xi' \sim |t'|^{-\nu}.
\end{equation} 

By dividing the two expressions, linearizing the transformation with a Taylor expansion~\cite{dombg}, taking the natural logarithm, and considering one renormalization group application, we obtain the following relation which is valid in the vicinity of the critical point $\beta_{c}$:
\begin{equation} \label{eq:cnu}
\nu = \frac{\log b}{\log \frac{d \beta'}{d \beta}}.
\end{equation}

Based on the above expression,  we are then able to calculate the correlation length exponent numerically. 

Due to the nature of disordered systems and, specifically, the averaging over disorder there exist two distinct approaches that can be utilized to calculate the correlation length exponent. In the former, conventional approach, we first conduct the thermal and disorder averaging to obtain $[\langle |q'|\rangle ]$, $[\langle |q| \rangle ]$ and then construct the inverse mappings:
\begin{equation} \label{eq:averag}
\beta'=[\langle |q| \rangle]^{-1}([\langle |q'|\rangle ](\beta)).
\end{equation}

The second approach concerns the construction of mappings between $\beta'$ and $\beta$ for individual realizations of disorder as:
\begin{equation}
\beta'_{\lbrace J_{ij}' \rbrace}=\langle |q_{\lbrace J_{ij} \rbrace}| \rangle^{-1}(\langle  |q_{\lbrace J_{ij}' \rbrace}'| \rangle( \beta_{\lbrace J_{ij} \rbrace} )).
\end{equation}

We remark that since, in the latter approach, one constructs mappings between individual realizations of disorder one expects significant fluctuations in the calculations. Consequently, to obtain a definite answer, one must reach lattice sizes for which the fluctuations of individual realizations of disorder are almost negligible: it therefore remains unclear if the later approach can be successfully studied even with the current state-of-the-art lattice sizes accessed by supercomputing facilities.    

To avoid constructing mappings between individual realizations of disorder, we consider for the results of this manuscript, only the former approach of Eq.~\ref{eq:averag}, namely we construct mappings after the thermal and disorder averaging of observables. To conduct the calculations we must determine the numerical derivative of Eq.~\ref{eq:cnu}. We therefore implement a fit using the nonlinear least-squares Marquardt-Levenberg algorithm. To evade an underrepresentation of the errors emerging from the fit range within the least-squares approach, we iteratively reduce the fit range and document the calculation with the highest error. Conclusively, we obtain for the correlation length $\nu=1.32(15)$.

\subsection{Explicit symmetry breaking and the exponent $\theta_{\epsilon}$}

We have so far utilized quantities derived from the overlap order parameter $[\langle |q| \rangle]$ to obtain independent calculations of the critical exponents $\nu$, $\gamma/\nu$ and $\kappa_{q}/\nu$. This is achieved by studying the corresponding phase transition based on the divergence of the correlation length $\xi$ in relation to the inverse temperature $\beta$.

We proceed to study a different setting with the Wilson two-lattice matching renormalization group. Specifically, we are now interested in coupling the overlap order parameter $q$ to a fictitious field $\epsilon$ and introducing it within the two-replica Hamiltonian of the three-dimensional Edwards-Anderson to induce explicit symmetry breaking. We will then construct mappings which relate the original and the renormalized fields $\epsilon$, $\epsilon'$ and we will extract an additional exponent, namely the exponent $\theta_{\epsilon}$ which, in this case, governs the divergence of the spin glass correlation length $\xi$ in relation to the fictitious field $\epsilon$.

To initiate our study we modify the Hamiltonian of~(\ref{eq:origham}) by introducing the overlap order parameter $q$ coupled to a fictitious field $\epsilon$. For simplicity, we will subsequently call $\epsilon$ the replica field. The modified Hamiltonian is:
\begin{equation}
E_{\sigma,\tau}'= E_{\sigma}+E_{\tau}-\epsilon Vq_{\sigma\tau},
\end{equation}
and defines a modified Boltzmann probability distribution:
\begin{equation}
p_{\sigma_{i},\tau_{j}}'= \frac{\exp[-\beta(E_{\sigma_{i}}+ E_{\tau_{j}}-\epsilon V q_{\sigma_{i}\tau_{j}})]}{\sum_{\sigma}\sum_{\tau} \exp[-\beta(E_{\sigma}+ E_{\tau}-\epsilon V q_{\sigma\tau})]}.
\end{equation}

We remark that, in principle, the above probability distribution could be directly sampled with Markov chain Monte Carlo simulations. Nevertheless we anticipate that, due to the introduced term, such simulations might be complicated. To avoid sampling we instead utilize reweighting to obtain expectation values pertinent to the above probability distribution based on the original configurations sampled with the probability distribution of~(\ref{eq:origprob}). Specifically, we consider, for a given realization of disorder, the expectation value of an arbitrary observable $O$ calculated under the two-replica probability distribution $p_{\sigma_{i},\tau_{j}}'$:
\begin{equation}
\langle O \rangle= \frac{\sum_{\sigma} \sum_{\tau} O_{\sigma\tau} \exp[-\beta(E_{\sigma}+ E_{\tau}-\epsilon V q_{\sigma\tau})]}{\sum_{\sigma}\sum_{\tau} \exp[-\beta(E_{\sigma}+ E_{\tau}-\epsilon V q_{\sigma\tau})]},
\end{equation} 
where the sums are over all configurations. We aim to approximate the above expectation value based on a numerical estimator with a set of sampling probabilities $\tilde{p}_{\sigma_{i},\tau_{j}}$:

\begin{equation}
\langle O \rangle= \frac{\sum_{i=1}^{N_{\sigma}} \sum_{j=1}^{N_{\tau}} O_{\sigma_{i}\tau_{j}} \tilde{p}_{\sigma_{i},\tau_{j}}^{-1} \exp[-\beta(E_{\sigma_{i}}+ E_{\tau_{j}}-\epsilon V q_{\sigma_{i}\tau_{j}})]}{\sum_{i=1}^{N_{\sigma}} \sum_{j=1}^{N_{\tau}} \tilde{p}_{\sigma_{i},\tau_{j}}^{-1} \exp[-\beta(E_{\sigma_{i}}+ E_{\tau_{j}}-\epsilon V q_{\sigma_{i}\tau_{j}})]}.
\end{equation} 

where $N_{\sigma}$ and $N_{\tau}$ correspond to the number of configurations sampled for each replica $\sigma$, $\tau$.

Since we will sample $p_{\sigma,\tau}'$ with the probability distribution of the original system we set $\tilde{p}_{\sigma,\tau}=p_{\sigma,\tau}$ and we obtain:
\begin{equation}
\langle O \rangle= \frac{\sum_{i=1}^{N_{\sigma}} \sum_{j=1}^{N_{\tau}} O_{\sigma_{i}\tau_{j}} \exp[\beta \epsilon V q_{\sigma_{i}\tau_{j}}]}{\sum_{i=1}^{N_{\sigma}} \sum_{j=1}^{N_{\tau}}  \exp[\beta \epsilon V q_{\sigma_{i}\tau_{j}}]}.
\end{equation} 

Using the above reweighting expression, which is agnostic to the original Hamiltonian~\cite{bachtis2020adding}, we are able to calculate expectation values of observables $O_{\sigma \tau}$ at inverse temperature $\beta$ for a nonzero replica field $\epsilon$. We achieve this by utilizing instead the original configurations at the same inverse temperature $\beta$ which have been simulated for a zero value of the replica field. We emphasize that, by substituting every occurence of $O$ with $O'$ we are additionally able to obtain expectation values for any renormalized observable $O'$.

\begin{figure}[t]
\begin{indented}
\item[]\includegraphics[width=8.6cm]{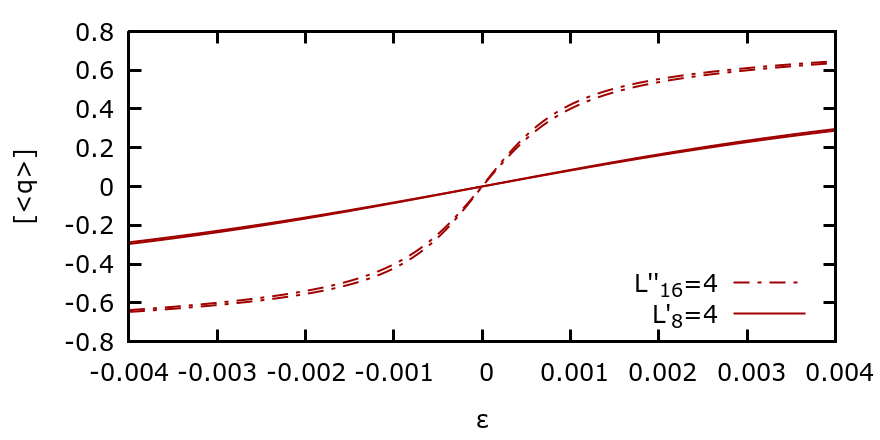}
\caption{\label{fig:figfield} Expectation of the overlap order parameter $q$ versus the replica field $\epsilon$ for two renormalized systems of identical lattice size. The space bounded by the lines corresponds to the statistical uncertainty.}
\end{indented}
\end{figure}

The expectation value of the overlap order parameter $[\langle q \rangle]$ for two renormalized systems of identical lattice size is depicted for nonzero values of the replica field $\epsilon$ in Fig.~\ref{fig:figfield}. We observe that the overlap order parameter $[\langle q \rangle]$  is driven towards positive or negative values when the replica field $\epsilon$ is positive or negative, respectively. This behavior is attributed to explicit symmetry breaking. Consequently, for positive or negative values of $\epsilon$ the spins of the two replicas tend to identical or exactly different values, respectively. 

We are now interested in utilizing the results depicted in Fig.~\ref{fig:figfield} to extract the critical exponent associated with the divergence of the spin glass correlation length $\xi$ in relation to the replica field $\epsilon$. To achieve this, we first construct inverse mappings which relate the original and the renormalized replica fields $\epsilon$, $\epsilon'$ via:
\begin{equation}
\epsilon'=O^{-1}(O'(\epsilon)).
\end{equation}   
\begin{figure}[t]
\begin{indented}
\item[]\includegraphics[width=8.6cm]{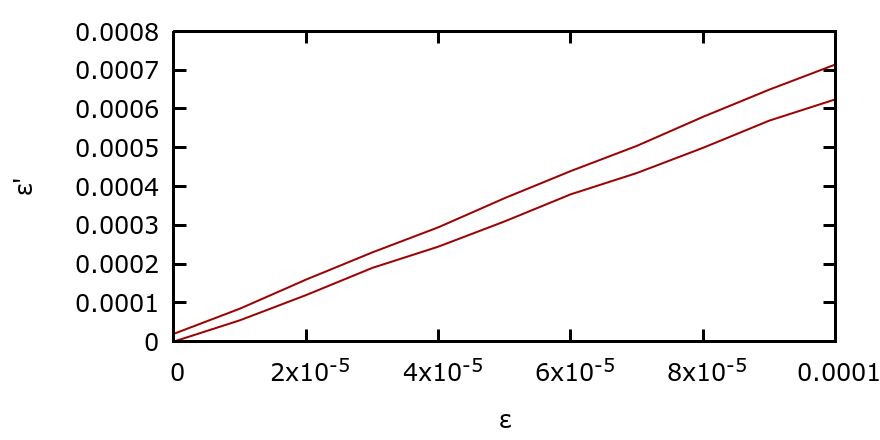}
\caption{\label{fig:figmapf} Renormalized replica field $\epsilon'$ versus original replica field $\epsilon$. The space bounded by the lines corresponds to the statistical uncertainty. }
\end{indented}
\end{figure}

 We now define an exponent $\theta_{\epsilon}$ which governs the divergence of the spin glass correlation length in terms of the replica fields $\epsilon$, $\epsilon'$ as:

\begin{equation}
\xi \sim |\epsilon|^{-\theta_{\epsilon}}, \ \xi' \sim |\epsilon'|^{-\theta_{\epsilon}}.
\end{equation}

The numerical calculation of the exponent $\theta$ is therefore achieved via:
\begin{equation}
\theta_{\epsilon}= \frac{\ln b}{\ln \frac{d \epsilon'}{d \epsilon}}.
\end{equation}

The above expression is valid in the vicinity of the critical fixed point $\beta_{c}$ and while $\epsilon \rightarrow 0$. Following the same approach as for the exponent $\nu$ we conduct the calculation of $\theta_{\epsilon}$ by constructing the mappings after the thermal and disorder average, which are depicted in Fig.~\ref{fig:figmapf}. We obtain a calculation of $\theta_{\epsilon}=0.369(8)$.

\subsection{Revisiting the calculations of $\gamma/\nu$, $\kappa_{q}/\nu$}

Having obtained a critical fixed point from the Wilson two-lattice matching Monte Carlo renormalization group method we can now conduct another set of calculations for the exponents $\gamma/\nu$, $\kappa_{q}/\nu$. The previous calculations of $\gamma/\nu$, $\kappa_{q}/\nu$ served only as a means to obtain an initial estimate of the fixed point without assuming any prior knowledge about the phase transition of the system. In fact, these calculations utilize equations which are strictly correct for the infinite-volume system and the estimates for these exponents should not be regarded as high-precision measurements.

 We will render the previous equations suitable for finite systems by linearizing the renormalization group transformation. In contrast to the previous calculations of $\gamma/\nu$, $\kappa_{q}/\nu$, which are conducted at a specific point in parameter space, we will now conduct the calculations via the use of finite differences. The benefit of considering a linearization in the vicinity of the phase transition is that one obtains robust calculations of critical quantities even if the system does not reside exactly at the critical fixed point.  Specifically, we utilize L'H$\mathrm{\hat{o}}$pital's rule to obtain~\cite{PhysRevLett.128.081603}:
\begin{equation}\label{eq:betaexp}
\frac{\kappa_{q}}{\nu}= \frac{\log \frac{dq'}{dq} }{\log b},
\end{equation}
\begin{equation}\label{eq:gammaexp}
\frac{\gamma}{\nu}= -\frac{\log \frac{d\chi_{q}'}{d\chi_{q}'} }{\log b}.
\end{equation}

\begin{figure}[t]
\begin{indented}
\item[]\includegraphics[width=8.6cm]{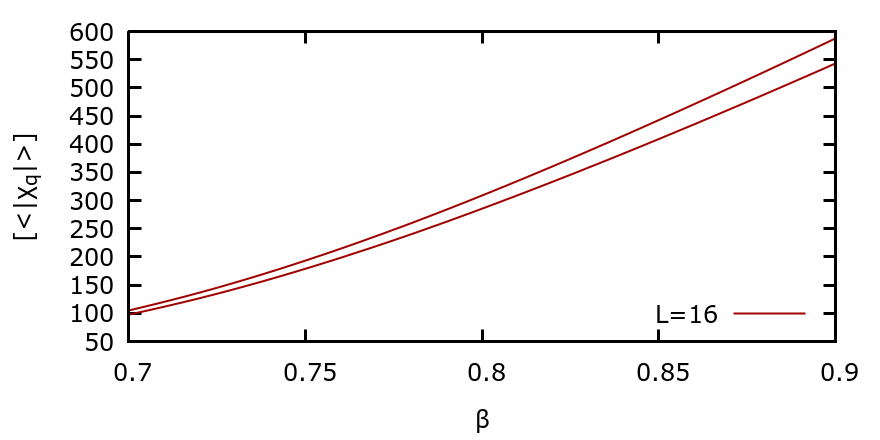}
\caption{\label{fig:figsusc} Expectation value of the overlap susceptibility $\chi_{q}$ versus the inverse temperature for lattice size $L=16$.  The space bounded by the lines corresponds to the statistical uncertainty. }
\end{indented}
\end{figure}

\begin{figure}[t]
\begin{indented}
\item[]\includegraphics[width=8.6cm]{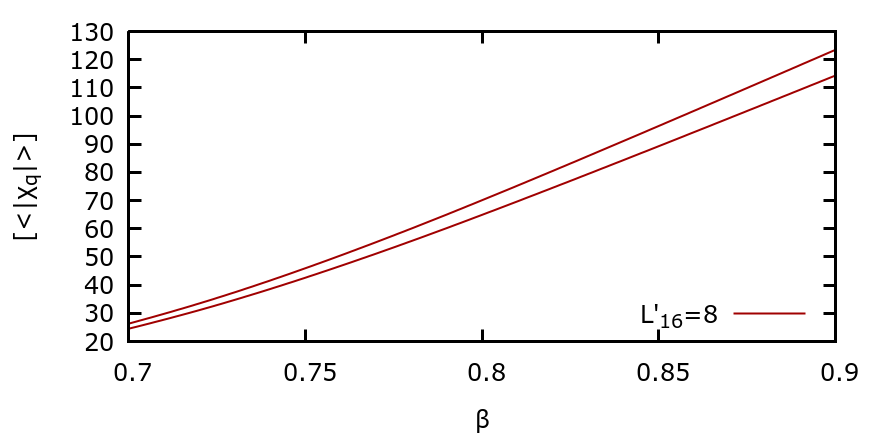}
\caption{\label{fig:figsusc2} Expectation value of the overlap susceptibility $\chi_{q}$ versus the inverse temperature for lattice size $L'_{16}=8$.  The space bounded by the lines corresponds to the statistical uncertainty. }
\end{indented}
\end{figure}

We now proceed to conduct the calculation of the exponents in the critical region which corresponds to the fixed point that emerged from the two-lattice matching approach. Based on the values of the overlap order parameter, which are depicted in Figs.~\ref{fig:figorig} and~\ref{fig:figlm}, we utilize systems with lattice sizes $\mathcal{L}'\neq \mathcal{L}$  to obtain a calculation of $\kappa_{q}/\nu=0.28(4)$. The values of the overlap susceptibility for the two systems are depicted in Figs.~\ref{fig:figsusc} and \ref{fig:figsusc2}. We obtain for the overlap susceptibility exponent $\gamma/\nu=2.30(5)$.

\section{Discussion}

The results of this manuscript, obtained via the Wilson two-lattice matching renormalization group method or from a direct computational implementation of the Kadanoff scaling picture are summarized in Table~\ref{tab:table2}. We find that calculations of the two critical exponents $\nu=1/{y_{t}}$ and $\theta_{\epsilon}=1/y_{h}$, which are obtained with the Wilson two-lattice matching method, are in agreement, within statistical errors, with Monte Carlo renormalization group calculations of the linearized matrix~\cite{PhysRevB.37.7745}. Based on the effective system, which comprises overlap degrees of freedom, it is conjectured within the Haake-Lewenstein-Wilkens~\cite{PhysRevLett.55.2606,PhysRevB.38.9086} perspective that the three-dimensional Edwards-Anderson model is characterized by a correlation length exponent $\nu$ which is two times the correlation length exponent of the conventional three-dimensional Ising model. The result of $\nu$, obtained on the given lattice sizes by first averaging over disorder and then constructing the mappings, supports this conjecture. Critical exponents for the three-dimensional Ising model have been calculated, for instance, with the conformal bootstrap method~\cite{Showk2014}.  We remark that different renormalization group transformations produce different fixed points~\cite{PhysRevLett.42.859}. Nevertheless, the emergence of a critical fixed point supports, based on renormalization group arguments, the evidence for the presence of a finite-temperature spin glass transition for the three-dimensional Edwards-Anderson model.

 Further potential research directions enabled by the Haake-Lewenstein-Wilkens approach could involve the construction of renormalization group mappings based on observables of the overlap variables, such as correlation functions $\varrho_{i} \varrho_{j}$, which are calculated under the disorder-averaged probability distribution. These could be utilized to determine the critical fixed point and to extract critical exponents. We remark that one anticipates, based on the Haake-Lewenstein-Wilkens approach, that a calculation of the correlation length exponent $\nu_{\textrm{eff}}$ based on correlation functions $\varrho_{i} \varrho_{j}$ would produce $\nu_{\textrm{eff}}=\frac{1}{2} \nu$~\cite{PhysRevB.37.7745}. One is then able to consider all possible correlation functions $\rho_{i} \rho_{j}$ on a given finite lattice to construct more accurate and systematically improvable renormalization group calculations which would be suitable for high precision measurements.

Alternative theoretical methods which are commonly used to obtain critical exponents and the critical inverse temperature  of systems that undergo phase transitions include finite-size scaling, nonequilibrium relaxation approaches and series expansions. These are generally conducted on the original Hamiltonian.  Critical exponents can additionally be calculated via experiments~\cite{PhysRevB.43.8199}. Specifically, $\mathrm{Fe_{0.5}Mn_{0.5}TiO_{3}}$ is anticipated to manifest behavior similar to the three-dimensional Edwards-Anderson model with the spins aligned along the hexagonal c-axis. The experimental  results are affected strongly by the choice of the critical temperature and are summarized in~\cite{young1998spin}. 

 Calculations of critical quantities~\cite{PhysRevLett.54.928,PhysRevB.32.7384,PhysRevB.31.340,PhysRevLett.57.245,PhysRevB.31.631,PhysRevLett.54.924,PhysRevB.37.5606,PhysRevB.53.R484,PhysRevLett.77.2798,Davidiguez_1996,PhysRevLett.80.4771,PhysRevB.58.14852,PhysRevLett.82.5128,PhysRevE.59.2653,PhysRevB.62.14237,PhysRevB.65.184409,TotaNakamura_2003,PhysRevB.72.184429,ParisenToldin_2006,PhysRevE.93.032126,PhysRevB.78.214205,PhysRevB.73.224432} are summarized, for instance, in~\cite{PhysRevB.73.224432}.   We remark that, within errors, calculations of the correlation length exponent in these theoretical studies range from $\nu=1.1$ to $\nu=3.9$: this is not surprising given the difficulty in studying spin glasses. Recent results,  which indicate a consistency in the values of the critical exponents, include calculations conducted on the Janus dedicated computer~\cite{PhysRevB.88.224416}. The calculations of the critical exponents $\beta/\nu$ and $\gamma/\nu$ are in agreement with the pertinent calculations of the Janus collaboration. It might be interesting to comment on the fact, discussed in~\cite{PhysRevB.73.224432} that, even within the same finite-size scaling analysis, the method utilized to obtain the infinite-volume limit extrapolation can produce substantially different values, such as $\nu=2.39(5)$ or $\nu=1.49(6)$.  

 We emphasize that, since four critical exponents have been calculated, it is possible to utilize scaling relations~\cite{newmanb99} to obtain multiple calculations of all exponents that describe the phase transition of the model. In fact, independent calculations of multiple critical exponents can additionally be utilized to verify if there exists a violation in scaling relations. For completeness, some scaling relations are:
\begin{equation}
\alpha= 2-\nu d,
\end{equation}
\begin{equation}
\frac{\kappa_{q}}{\nu}=d- \frac{1}{\theta_{\epsilon}},
\end{equation}
\begin{equation}
\frac{\gamma}{\nu}= \frac{2}{\theta_{\epsilon}}-d,
\end{equation}
\begin{equation}
\delta=\frac{1}{d \theta_{\epsilon} -1},
\end{equation}
\begin{equation}
\eta=d+2-\frac{2}{\theta_{\epsilon}},
\end{equation}
where $\alpha$ is the critical exponent which governs the singularity in the specific heat $c\sim |t|^{-\alpha}$, $\delta$ the exponent which governs the singularity of the overlap order parameter $q$ in terms of the replica field $\epsilon$, $q \sim \epsilon^{\frac{1}{\delta}}$, and $\eta$ the exponent defined from the two point correlation function $G$, $G(r,t) \sim 1/(r^{d-2+\eta})$. We have not assumed any form of correction or violation in the above equations.

Besides cross-verifying results with the four estimated critical exponents, it is of interest to utilize the estimations to calculate the critical exponent $\alpha$ of the specific heat. In general, to obtain a direct calculation of the exponent $\alpha$ within a renormalization group setting one would need to calculate the internal energy of the system and, consequently, one would need to renormalize the random couplings. Nevertheless, by utilizing the overlap order parameter $q$ to obtain an estimation of the correlation length exponent $\nu$ it is still possible to estimate $\alpha$, an exponent associated with an experimentally relevant quantity, without undergoing a renormalization of the random couplings. 

The results in this manuscript consider only statistical errors. Systematic errors that emerge in the calculations include the averaging over a finite number of realizations of disorder, the finite-size effects from considering small lattice sizes and the operators utilized to conduct the renormalization group calculation. More simulations on larger lattices are required.

 \begin{table}[t]
\begin{indented}
\item[] \begin{tabular}{ccccc}
\br 
&$\nu$ & $\kappa_{q}/\nu$ & $ \gamma/\nu$ & $\theta_{\epsilon}$ \\
\hline
W & $1.32(15)$ & $0.28(4)$  & $2.30(5)$ & $0.369(8)$ \\
K & - & $0.40(1)$ & $2.15(2)$  & - \\
\br
\end{tabular}
\caption{\label{tab:table2}
Estimates of the critical exponents $\nu$, $\kappa_{q}/\nu$, $\gamma/\nu$, $\theta_{q}$ for the three-dimensional Edwards-Anderson model. We denote the exponent of the overlap order parameter as $\kappa_{q}$ to avoid confusion with the inverse temperature $\beta$. The calculations are conducted based on the fixed points which emerged from the Wilson two-lattice matching approach (W) or the Kadanoff scaling picture (K).  }
\end{indented}

\end{table}
\section{Conclusions}

By mapping the spin glass phase transition of the three dimensional Edwards-Anderson model to an effective system, which manifests critical behavior that resembles ferromagnetism, we have shown that the implementation of renormalization group transformations on the overlap degrees of freedom enables the determination of the critical fixed point and the calculation of four critical exponents. Furthermore, we have conducted a renormalization group study of the explicit symmetry-breaking of the system. Specifically, we constructed inverse mappings to extract the function which relates the original and the renormalized field that is coupled to the overlap order parameter.

In addition, we have utilized reweighting techniques for renormalized observables under the original probability distribution to study the spin glass phase transition. These  would enable high-precision calculations for Monte Carlo renormalization group methods within the implementation of large-scale simulations. Specifically, when combined with parallel tempering, the multiple histogram method provides results in the entire critical region and simultaneously reduces the statistical errors: it is therefore possible to determine the critical fixed point via an intersection of original and renormalized observables without requiring additional simulations. Reweighting techniques in Monte Carlo renormalization group methods additionally overcome the need to simulate a Hamiltonian, which includes the overlap order parameter, to induce explicit symmetry breaking in the original or renormalized system. 

We reiterate that the current study is a proof-of-principle demonstration that is orders of magnitude smaller in computational effort than prior large-scale simulations or studies conducted with the use of supercomputing facilities, and it is not meant to compete in numerical precision with those. Potential benefits of Monte Carlo renormalization group methods include the partial elimination of finite-size effects~\cite{PhysRevB.37.7745}, the linearization of the transformation which enables accurate calculations of critical exponents anywhere in the linear region of the fixed point~\cite{PhysRevLett.42.859}, and the fact that configurations of smaller lattice sizes can be constructed by the application of the renormalization group transformation: all computational resources could then be employed to sample the largest lattice size possible within the context of a Monte Carlo renormalization group study and the smaller lattice sizes could be obtained via the application of the transformation and without requiring additional Monte Carlo simulations.

In summary, the utilization of renormalization group transformations on the overlap degrees of freedom enables the determination of the fixed point and the calculation of four critical exponents without the need to conduct a renormalization of the random couplings of the system. All other critical exponents can then be obtained via scaling relations. Consequently the discussed approach, enables studies of phase transitions that are fully characterized by overlap order parameters. It is therefore of interest to extend these techniques to other disordered systems, which also manifest glassy behavior.

\ack
The author thanks Giulio Biroli for discussions and comments on the manuscript and acknowledges support from the CFM-ENS Data Science Chair. 

%

%
%
%
%

\section*{References}

\bibliographystyle{iopart-num2}
\include{ms.bib}
\bibliography{ms}

\end{document}